# Room-temperature optically detected magnetic resonance of single defects in hexagonal boron nitride

Hannah L. Stern[1*†], John Jarman[1*], Qiushi Gu[1], Simone Eizagirre Barker[1], Noah Mendelson[2], Dipankar Chugh[3], Sam Schott[1], Hoe H. Tan[3], Henning Sirringhaus[1], Igor Aharonovich[2] and Mete Atatüre[1†].

[1] *Cavendish Laboratory, University of Cambridge, J.J. Thomson Ave., Cambridge, CB3 0HE, United Kingdom.*

[2] *ARC Centre of Excellence for Transformative Meta-Optical Systems, Faculty of Science, University of Technology Sydney, New South Wales, Australia.*

[3] *ARC Centre of Excellence for Transformative Meta-Optical Systems, Research School of Physics and Engineering, The Australian National University, Canberra, Australian Capital Territory, Australia.*

Optically addressable spins in materials are important platforms for quantum technologies, such as repeaters and sensors. Identification of such systems in two-dimensional (2d) layered materials offers advantages over their bulk counterparts, as their reduced dimensionality enables more feasible on-chip integration into devices. Here, we report optically detected magnetic resonance (ODMR) from previously identified carbon-related defects in 2d hexagonal boron nitride (hBN). We show that single-defect ODMR contrast can be as strong as 6% and displays a magnetic-field dependence with both positive or negative sign per defect. This bipolarity can shed light into low contrast reported recently for ensemble ODMR measurements for these defects. Further, the ODMR lineshape comprises a doublet resonance, suggesting either low zero-field splitting or hyperfine coupling. Our results offer a promising route towards realising a room-temperature spin-photon quantum interface in hexagonal boron nitride.

[*] These authors contributed equally to this work.

[†]Corresponding authors: H.L.S. (hs536@cam.ac.uk) and M.A. (ma424@cam.ac.uk).

# Introduction

Optically active confined spins in wide band-gap materials can act as 'artificial atoms' in convenient and scalable platforms[1,2]. Colour centres in diamond[3,4,5] and silicon carbide[6,7] are prime examples of such systems with long spin coherence times[8] and high-fidelity spin control and read-out[9]. Their coupling to nuclear spins further enables the realisation of optical accessible long-lived quantum memories [1,10-14]. Together with nanofabrication capabilities, these features make impurity spins leading candidates for quantum technologies[4,15-20].

Layered van der Waals materials are an alternative platform[21-25], where single-photon emitting defects are reported to be among the brightest to date[26] and the reduced dimensionality may allow for a feasible route to scalable quantum devices[27,28]. Hexagonal boron nitride (hBN) is a two-dimensional van der Waals crystal that was recently shown to host a plethora of defects that display photoluminescence (PL) spectra ranging from 580 nm to 800 nm[26,29,30]. These defects can also be tuned spectrally via strain and electric field [31-34]. Multiple defect classes are emerging in hBN: a structure involving a single negatively charged boron vacancy ($V_B^-$) displays broad emission at 800 nm and optically detected magnetic resonance (ODMR), as measured at the ensemble level [35-38]. There are also defects around 700 nm, where the presence of spin has been inferred via their magneto-optical signature[39], and recently via ODMR measurements on defect clusters[40]. A family of narrow-band bright emitters with distinctly sharper zero-phonon lines (ZPL) in the visible spectral range [41-43] has recently received more attention; they can be created controllably via chemical vapour deposition (CVD) [44-47] and plasma treatment methods[48], display spectrally narrow bright optical emission[49], and have already been integrated into optical cavities[50-52]. As such, they hold potential towards room-temperature devices for quantum-photonic applications and accessing their inherent spin at single-defect level is required for their implementation as a spin-photon interface.

In this Letter, we report room-temperature ODMR measurements on single hBN defects. We investigate hBN samples that host well-isolated single defects that have recently been assigned to carbon impurities and show that they present strong optical signatures of single spins. We find that the single-defect ODMR contrast can reach 6%, approximately 15-fold stronger signal over the 0.4% contrast we observe for the ensemble. Moreover, the bipolar nature of the ODMR contrast sign likely contributes to the low contrast of the ensemble. Finally, high-resolution ODMR lineshape measurements indicate a double-peaked resonance for each defect that may arise from finite zero-field splitting or hyperfine coupling to proximal or host nuclear spins.

## Results

*Material Characterisation.*

To compare the behaviour of ensembles of hBN defects with the behaviour of singles, we measure a series of hBN samples with varying defect density, grown via MOVPE using a carbon rich precursor, where the optical emission has been associated with carbon impurities[47] (see Methods). The samples show increasing levels of carbon-boron and carbon-nitrogen bonding which in turn correlates with the defect density and brightness of the material under 532-nm illumination[47].

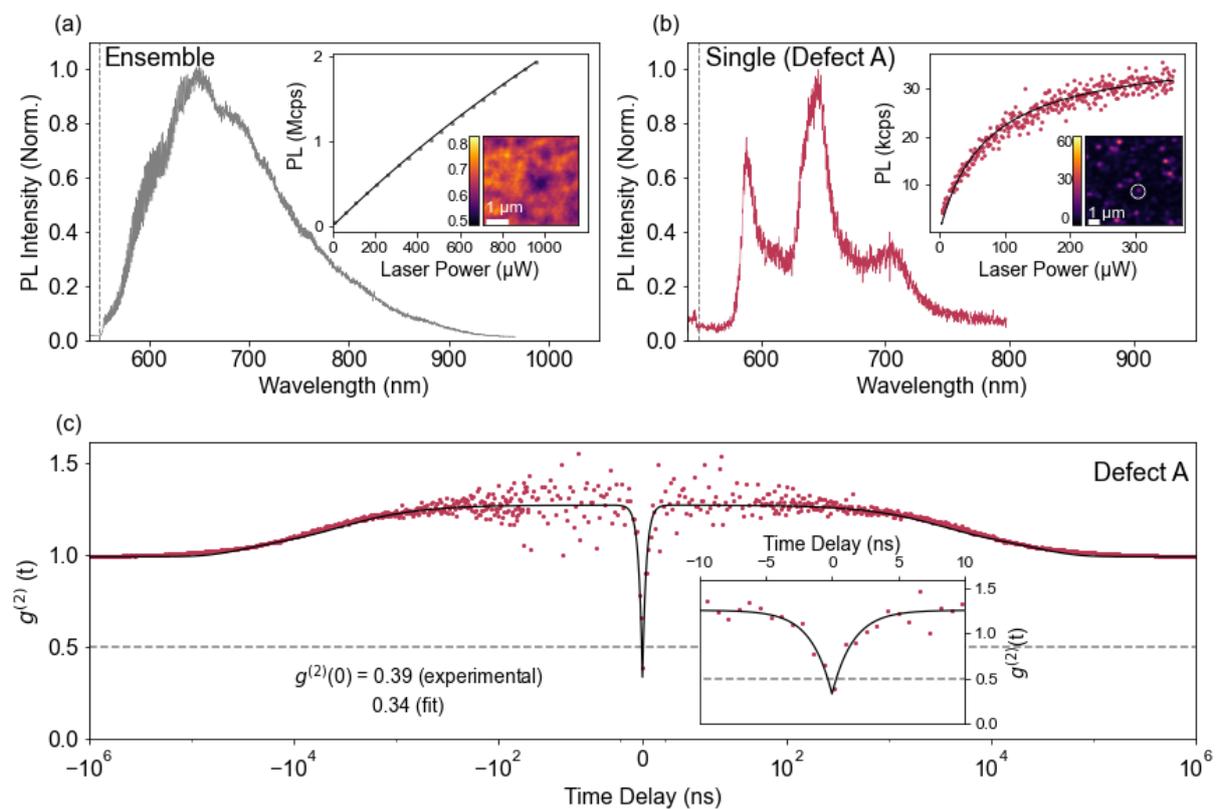

**Figure 1: Optical Properties of single and ensemble hBN Defects.** (a) Normalised PL spectrum of ensemble defects under 532-nm laser excitation. Dashed vertical line represents the cut-off of a 550-nm long pass filter. Inset: 5x5-µm image map of the integrated PL intensity and example laser power dependence of PL counts at a fixed location. The colour bar scale is in Mcps. (b) Normalised PL spectrum of single defect A under 532-nm laser excitation. Dashed vertical line signifies the 550-nm long pass filter. Inset: 10x10-µm image map of the integrated PL intensity with a white circle around defect A, and the power dependence for defect A, showing $P_{sat}^{optical}$ = 70(2) µW and and Isat = 37,900(300) counts/s. (c) Second-order correlation measurement for defect A at 1.4$P_{sat}^{optical}$ excitation, solid black curve is a theoretical fit. Inset: A zoom of the same measurement for short timescales revealing the fitted $g^{(2)}(0)$ = 0.34 and optical lifetime of 1.60(1) ns.

Figure 1 presents representative optical properties for the defect ensemble in the high-density sample (panel a) and for defect A, a typical isolated defect in the low-density sample (panels b and c). The insets of the panels a and b include integrated-PL intensity as a function of optical excitation power, as well as integrated-PL confocal images showing the defect density for the two samples. In contrast to the broad PL spectrum for the ensemble (Fig. 1a), the single-defect spectrum in Fig. 1b comprises well-resolved ZPL and multiple phonon sidebands (PSB) with an energy tuning of ~170-180 meV, consistent with previous reports[47]. Figure 1c is the second-order autocorrelation measurement $g^{(2)}(\tau)$ on the integrated-PL intensity for defect A. In addition to the antibunching behaviour of $g^{(2)}(0) = 0.34$ (Fig. 1c inset) confirming that defect A is an isolated single defect, we observe clear bunching at microsecond timescales, consistent with previous reports, and all other defects we investigated (see SI).

*Optically Detected Magnetic Resonance.*

Figure 2a illustrates the basic elements of the continuous wave ODMR setup. We record integrated-PL intensity under 532-nm laser excitation as a function of the applied microwave field. We modulate the amplitude of the microwave field with a square wave at 70 Hz to determine the difference between the PL intensity when the microwave field is applied (signal) and when the field is not present (reference). The difference in PL is normalised by the reference PL intensity to obtain an ODMR contrast for each microwave frequency. This eliminates contributions from slow variations in the course of each measurement. A permanent magnet on a linear stage tunes the amplitude of the external magnetic field at the defect, which is applied in-plane relative to the sample. Figure 2b shows an optical image of one of our hBN devices, where we rely on lithographically patterned microstrip on the hBN layer to deliver the microwave field locally. The microstrip is deposited on top of the grown hBN layer, which uniformly spans the whole image.

Figure 2c presents an example ODMR spectra for the ensemble (grey circles) and two single defects (red and orange circles) at 25-mT in-plane magnetic field. All three saturated ODMR signals are at 700-MHz central frequency and ~35-MHz linewidth. Strikingly, the single-defect ODMR signals have substantially higher (more than 15-fold) contrast with respect to that of the ensemble. The comparable linewidth observed for the ODMR spectrum of the high-density ensemble and the single defects suggests that the mismatch might arise from a possibly low fraction of spin-active defects, similar to previous reports[39], as opposed to other effects such as spectral broadening. Indeed, out of 80 isolated single defects investigated within a 40 x 40 um area, only nine revealed measurable ODMR signal suggesting a yield of ~10%. Further, defects A and B in Fig. 2c are examples of the positive and negative ODMR contrast that we observe across the ODMR active defects, with a roughly even yield of each polarity (see SI). A positive (negative) ODMR signal indicates that microwave drive at spin resonance frequency leads to an increased (decreased) PL intensity, which can further contribute to the modest ODMR signal from the ensemble. Figure 2d presents the ODMR contrast of the single defect A as a function of microwave power at 25-mT applied magnetic field, demonstrating the expected saturation behaviour. The same measurement for the ensemble ODMR contrast shows equivalent saturation behaviour albeit at a significantly lower ODMR signal.

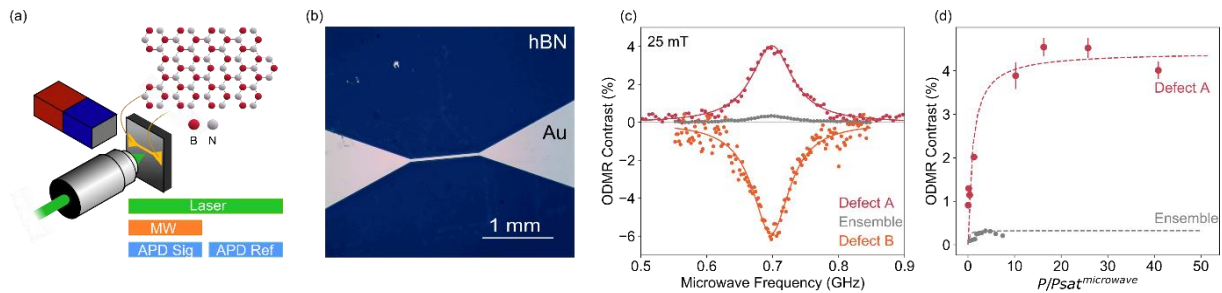

**Figure 2: Room Temperature ODMR Setup and Measurements.** (a) An illustration of the measurement set-up showing a permanent fixed magnet positioned in-plane relative to the hBN sample with lithographically patterned microstrip. Schematic representing one cycle of the ODMR protocol: a microwave pulse (orange) for the first half of the lock-in cycle, and signal and reference counts that are measured by the single-photon counting detectors (APD) (blue). The excitation laser is present for the full lock-in cycle (green). (b) An optical image of a lithographically patterned 1-mm-long gold microstrip used to apply microwave field to the hBN defects (see SI). (c) Room-temperature ODMR of single defects (defect A, red circles; defect B, orange circles) and of the ensemble (grey circles), all measured at 25-mT external magnetic field and at $10 P_{sat}^{microwave}$. The solid curves are Lorentzian fits to the ODMR lineshapes. We determine a linewidth of 34(3), 37(2) and 34(2) for the ensemble, defect A and defect B, respectively. (d) ODMR contrast as a function of normalised microwave power ($P/P_{sat}^{microwave}$) for defect A (red) and the ensemble (grey).

*Magnetic-field dependence of single-defect ODMR.*

An ODMR frequency of 700 MHz at 25 mT is consistent with a g-factor of ~2, typical for atomic spin defects in solids and Fig. 3 presents the evolution of the ODMR spectra for single defects A (top row), B (middle row) and C (bottom row) up to an in-plane magnetic field of 140 mT, with the corresponding PL spectra shown in Fig. 3a for comparison. The ODMR spectra in Fig. 3b are all acquired at a fixed input microwave power ($10 P_{sat}^{microwave}$ at 25 mT for defect A), in order to compromise between microwave-induced heating at high field and signal strength at low field. The apparent variation of contrast, common to all defects, is due to the frequency-dependent microwave transmission into the microstrip. The black crosses highlight the saturated ODMR contrast for the corresponding spectra for defect A (see SI), which shows that the maximum ODMR contrast builds up to a constant value of 4% as a function of the magnetic field. Figure 3c presents the magnetic-field-dependent shift of the central frequency for the ODMR signal for these defects. A linear fit to each plot reveals a g-factor of 1.98(1), 1.97(1), and 1.98(1) for defects A, B and C, respectively, all in line with the measured value of 2.03(1) for the ensemble (see SI).

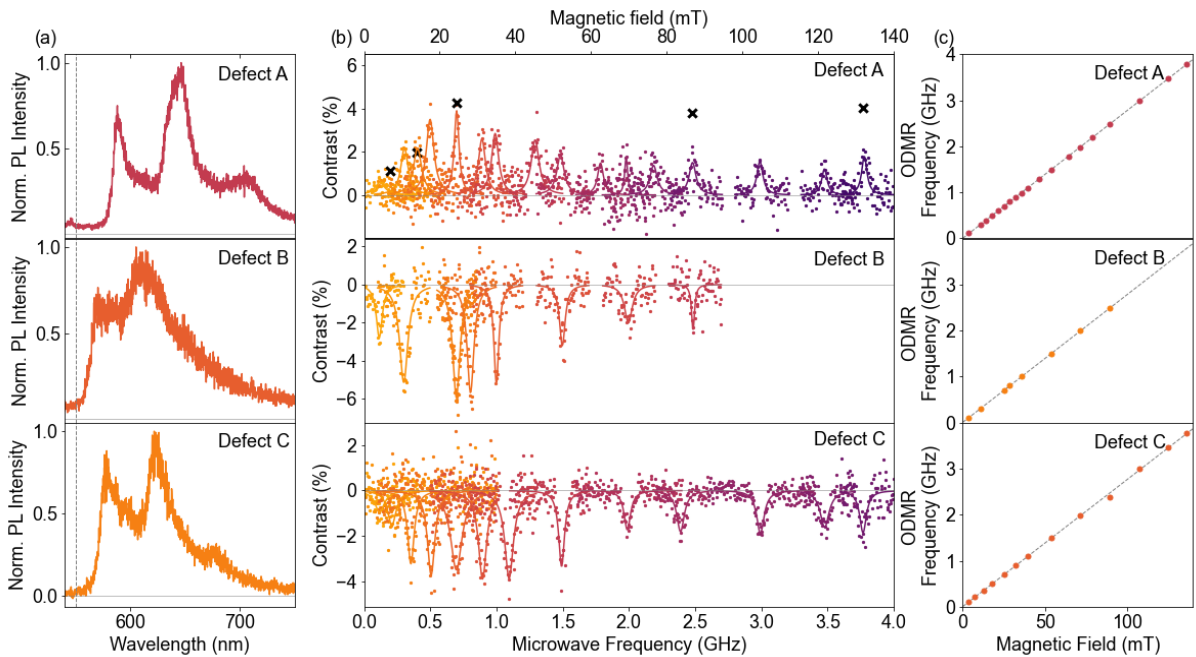

**Figure 3: Magnetic-Field Dependence of ODMR Resonance.** (a) Integrated PL spectra of defects A, B and C under 532-nm optical excitation. (b) ODMR spectra for defects A, B and C as a function of in-plane magnetic field, measured at 0.2W, which is $10 P_{sat}^{microwave}$ at 25 mT and $2 P_{sat}^{microwave}$ at 89 mT. A constant microwave power was used across the magnetic field range as higher powers cause microwave-induced heating. The black crosses mark the saturated contrast at that magnetic field strength, i.e., 1.1 % and 1.9% at 7 mT and 14 mT, respectively, saturating at ~4% at 25 mT and beyond (see SI). (c) ODMR resonance central frequency against magnetic field, fit to a linear function. The g-factors extracted from the linear fits are 1.98(1), 1.97(1) and 1.98 (1) for defects A, B and C, respectively.

To resolve sub-linewidth features in the ODMR spectrum, we operate in the near-optimal regime of signal strength with both optical and microwave excitation power at $0.75 P_{sat}^{optical}$ and $P_{sat}^{microwave}$, as inferred from saturation measurements (see SI). Figure 4 presents the corresponding ODMR spectra obtained at a magnetic field of 25 mT for the single defects A, B and C, as displayed in the panels a, b and c, respectively. The measurements (coloured circles) are fit with two separate functions for comparison. In each panel, the solid curve (red to orange) is a Lorentzian doublet, while the dashed grey curve is a single Lorentzian function best fit (for Gaussian and Voigt fit attempts, see SI). Each lower sub-panel displays the residuals for the doublet (coloured red to orange) and single (grey) Lorentzian fits to the ODMR spectrum. For all three defects the best fit is the double Lorentzian and yields an ODMR central-frequency splitting of 29(2), 35(2) and 42(3) MHz, for defect A, B, and C, respectively.

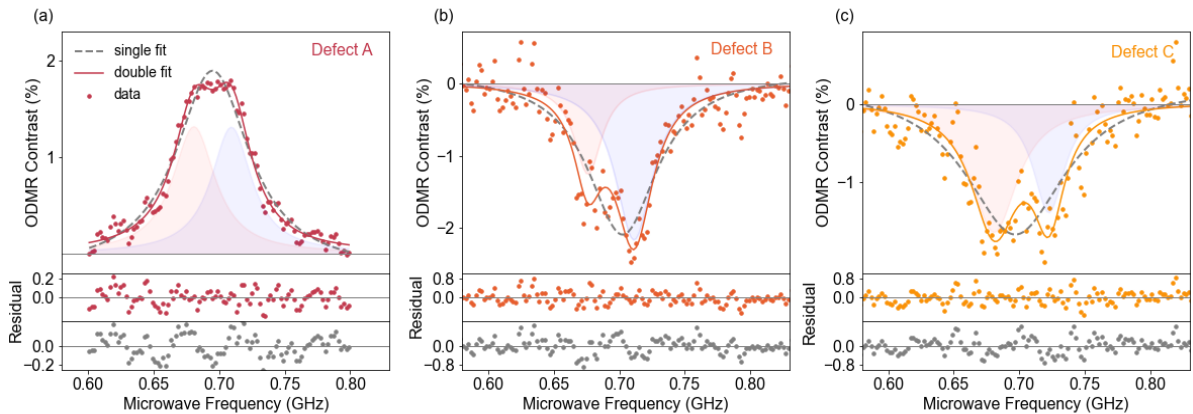

**Figure 4: Sub-Linewidth Structure of ODMR Spectrum.** (a-c) ODMR spectra with single and double Lorentzian fits for defects A, B and C, respectively. Coloured circles are the experimental results obtained at 25 mT, coloured solid curves are double-Lorentzian fits, grey dashed curves are single-Lorentzian fits for reference. For each panel, the shaded blue and red lineshapes show the two components of the double-Lorentzian fit. Residuals for double Lorentzian (coloured circles) and single (grey circles) Lorentzian fits are presented below the corresponding ODMR spectrum.

## Discussion

For the defects that show ODMR, the behaviour we observe is remarkably similar across the single defects we studied with roughly equal likelihood on the sign of the signal. This polarity is unlike the defects in diamond and SiC that show a consistent ODMR sign for a given optical defect whether probed as a single or on the ensemble level. In SiC the 4H and 6H polytypes contain multiple optical defects that have spectrally indistinguishable ZPLs at room temperature, yet display zero-field ODMR contrast at different microwave frequencies and with different sign[53,54]. For the two defects shown in Fig. 2 (and an additional shown in SI) we find little correlation between ZPL position and sign of ODMR, but this analysis is convoluted by the spectrally broadened ZPL lineshape of hBN defects at room temperature[49]. It is possible we have defects with very similar optical spectra but different ODMR behaviour, as in the case of SiC. However, unlike SiC, the negative and positive contrast defects all possess the same g-factor of ~2. An alternative explanation is that the presence of local strain could be influencing the optical emission properties of each defect, and the coexistence of negative and positive ODMR reflects an intricate balance of the relevant optical selection rules dictating spin initialisation and readout. Without further insight into the atomistic nature of the defects, it is difficult to distinguish the two possibilities and it is nevertheless plausible that this bipolar ODMR contrast contributes to the substantially lower ODMR contrast for the ensemble.

Both zero-field splitting in the high-field regime and hyperfine coupling can in principle lead to the split doublet in the ODMR spectrum. Electron paramagnetic resonance (EPR) measurements have shown that electronic spins in hBN couple to nitrogen, carbon and boron nuclear spins [35,55,56]. However, the predicted hyperfine constants and the corresponding splitting for boron[55] and carbon[40] isotopes differ starkly from our results. The hyperfine constant for nitrogen is in the correct range[35], but we do not expect a doublet spectrum from electron-nitrogen coupling. Bright and dark EPR measurements at 335 mT on the high-density sample reveal a single light-enhanced EPR absorption with 20-MHz linewidth (see SI), broadly in agreement with the ensemble ODMR spectrum, and do not show features that would suggest hyperfine coupling at the predicted coupling constants. The alternative explanation for the ODMR doublet is low zero-field splitting of a spin >½ state, although we note that the magnitude of potential zero-field splitting would be much smaller than other similar systems[5,53,57]. Therefore, we refrain from assigning the spin multiplicity and the exact nature of the ODMR structure.

## Conclusions

In conclusion, we report optically accessible spin defects in hBN layers via ODMR measurements at room temperature. We consistently observe 4-6% ODMR contrast for single well-isolated defects. The sub-unity yield of the ODMR-displaying defects, as well as the polarity of the ODMR sign, are likely reasons for the significantly reduced ODMR contrast reported previously for an ensemble. All ODMR-active defects possess a double peaked resonance with an average splitting of 35 MHz, suggesting a S>½ state with a zero-field splitting on the order of 20 MHz, though we are not ruling out hyperfine effects due to host or proximal nuclei. Further experimental and theoretical work will be required to develop a deeper insight into the microscopic photophysics of these defects. That said, these results underline the potential for these defects as a room-temperature spin-photon interface in a two-dimensional material platform.

## Methods

**HBN Layers.** hBN was grown by metal organic vapor phase epitaxy (MOVPE) on sapphire, as described in detail previously[47]. Briefly, triethyl boron (TEB) and ammonia were used as boron and nitrogen sources with hydrogen used as a carrier gas. Growth was performed at low pressure (85 mBar) and at a temperature of 1350 °C. Isolated defects and ensemble defects were generated by modifying the flow rate of TEB during growth, a parameter known to control the incorporation of carbon within the resulting hBN film. For PL measurements, hBN films were transferred to $SiO_2$/Si substrates, using a water-assisted self-delamination process to avoid polymer contamination. Before measurements, each sample was treated in a UV/ozone cleaner for 15 minutes.

**Confocal Microscopy.** Optical measurements were carried out at room temperature under ambient conditions using a home-built confocal microscopy setup. A continuous-wave 532-nm laser (Ventus 532, Laser Quantum) was sent through a 532 nm band-pass filter, and focused on the device using an objective lens with 100x magnification and a numerical aperture of 0.9. Control over excitation power was provided by an acousto-optic modulator (AA Optoelectronics), with the first-order diffracted beam fibre-coupled into the confocal setup. Two 550 nm long-pass filters (Thorlabs FEL550) were used to filter off reflected laser light from the collected emission, which was then sent either into an avalanche photodiode (APD) (SPCM-AQRH-14-FC, Excelitas Technologies) for recording photon count traces and observing the intensity of emission, or to a CCD-coupled spectrometer (Acton Spectrograph, Princeton Instruments) via single-mode optical fibres (SM450 and SM600) for photoluminescence spectroscopy measurements. White-light images of the device were collected by introducing the flippable mirror to divert the collection path to a CCD instead of the detection arm. This allowed easy device positioning and identification. Correlation measurements were carried out using a Hanbury Brown and Twiss interferometry setup using a time-to-digital converter (quTAU, qutools) with 80 ps resolution.

**Optically Detected Magnetic Resonance Measurements.** ODMR measurements were performed on the confocal setup described above. A 20-μm microstrip microwave antenna was patterned using standard photolithography techniques and thermal evaporation of 100 nm Au on 20 nm Ti. The antenna was bonded to a coplanar waveguide on a printed circuit board (PCB), shorting the waveguide at the sample. A 70-Hz square-wave modulation was applied to the microwave amplitude to detect the change in PL counts as a function of microwave frequency. A permanent magnet delivers an external static magnetic field in the plane of the hBN surface and is changed in strength and orientation by displacing the magnet.

**Electron Paramagnetic Resonance Measurements.** CW-EPR measurements were performed on TEB30 using a Bruker E500 X-band spectrometer with a ER 4122SHQE cavity at a microwave frequency of 9.370 GHz and a microwave power of 2mW. The external magnetic field was modulated at 100 kHz with an amplitude of 0.5 mT and the spectra were recorded as first harmonics of the absorption signal. The sample was mounted in the center of the cavity on a high-purity quartz slide. 532 nm laser light was coupled into the cavity through an optical window and focused on to the sample with a spot size of ~ 2.5 mm diameter. All ESR measurements were carried out at room temperature as a function of the average laser power ranging from 10 mW to 130 mW. They are reported after background subtraction from the sample mount.

## Acknowledgements

We would like to thank L. Bassett, A. Alkauskas and S. Stranks for useful discussions. We thank the research group of Sam Stranks for use of their laser for the EPR measurements. This work is funded by ERC Proof-of-Concept grant PEGASOS (862405), ERC Advanced Grant PEDESTAL (884745), Australian Research Council (CE200100010), and the Asian Office of Aerospace Research & Development (FA2386-20-1-4014). H.L.S is funded by Trinity College, Cambridge. S.E.B. acknowledges funding from the EPSRC CDT in Nanoscience and Nanotechnology (NanoDTC, Grant No. EP/S022953/1). Q.G. acknowledges financial support by the China Scholarship Council and the Cambridge Commonwealth, European & International Trust. S.S. acknowledges funding from the ERC Synergy Grant SC2 No. 610115.

## References

1. Awschalom, D. D., Hanson, R., Wrachtrup, J. & Zhou, B. B. Quantum technologies with optically interfaced solid-state spins. *Nat. Photonics.* **12**, 516–527 (2018).
2. Atatüre, M., Englund, D., Vamivakas, N., Lee, S.-Y. & Wrachtrup, J. Material platforms for spin-based photonic quantum technologies. *Nat. Rev. Mater* **3**, 38–51 (2018).
3. Gruber, A. *et al.* Scanning confocal optical microscopy and magnetic resonance on single defect centres. *Science.* **276**, 2012–2014 (1997).
4. Acosta, V. & Hemmer, P. Nitrogen-vacancy centers: physics and applications. *MRS Bull.* **38**, 127–130 (2013).
5. Doherty, M. W. *et al.* The nitrogen-vacancy colour centre in diamond. *Phys. Rep.* **528**, 1–45 (2013).
6. Christle, D. J. *et al.* Isolated electron spins in silicon carbide with millisecond coherence times. *Nat. Mater.* **14**, 160–163 (2014).
7. Seo, H., Falk, A. L., Klimov, P. V., Miao, K. C., Galli, G., and Awschalom, D. D. Quantum decoherence dynamics of divacancy spins in silicon carbide. *Nat. Commun.* **7**, 12935 (2016).
8. Maurer, P. C. *et al.* Room-temperature quantum bit memory exceeding one second. *Science.* **336**, 1283–1286 (2012).
9. Robledo, L., Childress, L., Bernien, H. *et al.* High-fidelity projective read-out of a solid-state spin quantum register. *Nature.* **477**, 574–578 (2011).

10. Pla, J.J. *et al.* High-fidelity readout and control of a nuclear spin qubit in silicon. *Nature.* **496**, 334 (2013).

11. Childress L., Gurudev Dutt M.V., Taylor J.M., Zibrov A.S., Jelezko F., Wrachtrup J., Hemmer P.R., Lukin M.D. Coherent dynamics of coupled electron and nuclear spin qubits in diamond. *Science.* **13**, 314, (2006).

12. Bradley, C.E., Randall, J., Abobeih, M.H., Berrevoets, R.C., Degen, M.J., Bakker, M.A., Markham, M., Twitchen, D.J., Taminiau, T.H., A ten-qubit solid-state spin register with quantum memory up to one minute, *Phys. Rev. X*, **9**, 3, (2019).

13. Sukachev, D. D., Sipahigil, A., Nguyen, C. T., Bhaskar, M. K., Evans, R. E., Jelezko, F., and Lukin, M. D. Silicon-vacancy spin qubit in diamond: a quantum memory exceeding 10 ms with single-shot state readout. *Phys. Rev. Lett.* **119**, 223602 (2017).

14. Metsch, M. H. *et al.* Initialization and readout of nuclear spins via a negatively charged silicon-vacancy centre in diamond. *Phys. Rev. Lett.* **122**, 190503 (2019).

15. Aharonovich, I., Englund, D. & Toth, M. Solid-state single-photon emitters. *Nature Photon* **10**, 631–641 (2016).

16. Sipahigil A. *et al.* An integrated diamond nanophotonics platform for quantum-optical networks. *Nature*, **354**, 6314 (2016).

17. Evans, R. E. *et al.* Photon-mediated interactions between quantum emitters in a diamond nanocavity. *Science* **362**, 662–665 (2018).

18. Taminiau, T. H., Cramer, J., van der Sar, T., Dobrovitski, V. V. & Hanson, R. Universal control and error correction in multi-qubit spin registers in diamond. *Nat. Nanotech.* **9**, 171–176 (2014).

19. Maze, J., Stanwix, P., Hodges, J. *et al.* Nanoscale magnetic sensing with an individual electronic spin in diamond. *Nature* **455**, 644–647 (2008).

20. Schmitt *et al.* Submillihertz magnetic spectroscopy performed with a nanoscale quantum sensor. *Science*, **356**, 6340 (2017).

21. Srivastava, A., Sidler, M., Allain, A. V., Lembke, D. S., Kis, A. and Imamoglu, A. Optically active quantum dots in monolayer $WSe_2$. *Nat. Nanotech.* **10**, 491–496 (2015).

22. He, Y. M. *et al.* Single quantum emitters in monolayer semiconductors. *Nat. Nanotech.* **10**, 497–502 (2015).

23. Koperski, M. *et al.* Single photon emitters in exfoliated WSe2 structures. *Nat. Nanotech.* **10**, 503–506 (2015).

24. Chakraborty, C., Kinnischtzke, L., Goodfellow, K. M., Beams R. and Vamivakas, A. N. Voltage-controlled quantum light from an atomically thin semiconductor *Nat. Nanotech.* **10**, 507–511 (2015).

25. Tonndorf, P. *et al.* Single-photon emission from localized excitons in an atomically thin semiconductor. *Optica.* **2**, 347–352 (2015).

26. Tran, T. T., Bray, K., Ford, M. J., Toth, M. & Aharonovich, I. Quantum emission from hexagonal boron nitride monolayers. *Nat. Nanotechnol.* **11**, 37–41 (2016).

27. Palacios-Berraquero, C. *et al.* Large-scale quantum-emitter arrays in atomically thin semiconductors. *Nat Commun.* **8**, 15093 (2017).

28. Branny, A., Kumar, S., Proux, R., and Gerardot, B. D. Deterministic strain-induced arrays of quantum emitters in a two-dimensional semiconductor. *Nat. Commun.* **8**, 15053 (2017).

29. Jungwirth, N.R, Calderon, B., Ji, Y., Spencer, M. G., Flatte, M. E., and Fuchs, G. Temperature dependence of wavelengths selectable zero-phonon emission from single defects in hexagonal-boron nitride. *Nano Lett.* **16**, 10, 6052–6057 (2016).

30. Tran, T.T., Elbadawi, C., Totonjian, D., Lobo, C.J., Grosso, G., Moon, H., Englund, D. R., Ford, M.J., Aharonovich, I., and Toth, M. Robust multicolor single photon emission from point defects in hexagonal boron nitride. *ACS Nano* **10**, 8, 7331–7338 (2016).

31. Grosso, G. *et al.* Tunable and high-purity room temperature single-photon emission from atomic defects in hexagonal boron nitride. *Nat Commun.* **8**, 705 (2017).

32. Noh, G. *et al.* Stark tuning of single-photon emitters in hexagonal boron nitride. *Nano Lett.* **18**, 4710–4715 (2018).

33. Proscia, N. V. *et al.* Near-deterministic activation of room temperature quantum emitters in hexagonal boron nitride. *Optica* **5**, 1128–1134 (2018).

34. Nikolay, N. *et al.* Very large and reversible Stark-shift tuning of single emitters in layered hexagonal boron nitride. *Phys. Rev. Appl.* **11**, 041001 (2019).

35. Gottscholl, A. *et al.* Initialization and read-out of intrinsic spin defects in a van der Waals crystal at room temperature. *Nat. Mater.* **19**, 540–545 (2020).

36. Ivády, V., Barcza, G., Thiering, G., Li, S., Hamdi, H., Chou, J., Legeza, O., and Gali, A. Ab initio theory of the negatively charged boron vacancy qubit in hexagonal boron nitride. *npj Comput. Mater.* **6**, 41 (2020).

37. Gottscholl, A., Diez, M., Soltamov, V., Kasper, C., Sperlich, A., Kianinia, M., Bradac, C., Aharonovich, I. and Dyakonov, V. Room temperature coherent control of spin defects in hexagonal boron nitride. Arxiv preprint: arXiv:2010.12513 (2020).

38. Liu, W. *et al*. Rabi Oscillation of VB- spin in hexagonal boron nitride. Arxiv preprint: arXiv:2101.11220 (2021).

39. Exarhos, A.L. *et al.* Magnetic-field-dependent quantum emission in hexagonal boron nitride at room temperature. Nat. Commun. **10**, 222 (2019).

40. Chejanovsky, N. *et al.* Single spin resonance in van der Waals embedded paramagnetic defect. Arxiv preprint: 1906.05903 (2019).

41. Exarhos, A. L., Hopper, D. A., Grote, R. R., Alkauskas, A. & Bassett, L. C. Optical signatures of quantum emitters in suspended hexagonal boron nitride. *ACS Nano.* **11**, 3328–3336 (2017).

42. Shotan, Z. *et al.* Photoinduced modification of single-photon emitters in hexagonal boron nitride. *ACS Photonics* **3**,12, 2490-2496 (2016).

43. Khatri, P., Ramsay, A.J., Malein, R. N. E., Chong, H. M. H., and Luxmoore, I. J. Optical gating of photoluminescence from colour centers in hexagonal boron nitride. *Nano Lett.* **20**, 6, 4256–4263 (2020).

44. Mendelson, N. *et al.* Engineering and tuning of quantum emitters in few-layer hexagonal boron nitride. *ACS Nano.* **13**, 3, 3132–3140 (2019).

45. Abidi, I.H. *et al.* Selective defect formation in hexagonal boron nitride. *Adv. Opt. Mater.* **7**, 1900397 (2019).

46. Stern, H.L. *et al.* Spectrally resolved photodynamics of individual emitters in large-area monolayers of hexagonal boron nitride. *ACS Nano.* **13**, 4538-4547 (2019).

47. Mendelson, N., Chugh, D., Reimers, J.R. *et al.* Identifying carbon as the source of visible single-photon emission from hexagonal boron nitride. Nat. Mater. **20**, 321–328 (2021).

48. Vogl, T., Campbell, G., Buchler, B. C., Lu, Y. & Lam, P. K. Fabrication and deterministic transfer of high-quality quantum emitters in hexagonal boron nitride. *ACS Photonics.* **5**, 2305–2312 (2018).

49. Dietrich, A., Doherty M. W., Aharonovich I., Kubanek A. Solid-state single photon source with Fourier transform limited lines at room temperature. *Phys. Rev. B,* **101**, 081401 (2020).

50. Vogl, T., Lecamwasam, R., Buchler, B. C., Lu Y., and Lam, P.K. Compact cavity-enhanced single-photon generation with hexagonal boron nitride. *ACS Photonics*, **6**, 8, 1955–1962 (2019).

51. Proscia, N.V., Jayakumar, H., Ge, X., Lopez-Morales, G., Shotan, Z., Zhou, W., Meriles, C. A. and Menon, V. M. Microcavity-coupled emitters in hexagonal boron nitride. *Nanophotonics*, **9**, 2937-2944 (2020).

52. Froch, J., Kim, S., Mendelson. N., Kianinia. M., Toth. M. and Aharonovich, I. Coupling hexagonal boron nitride quantum emitters to photonic crystal cavities. *ACS Nano.* 14, 6, 7085–7091 (2020).

53. Koehl, W., Buckley, B., Heremans, F. *et al.* Room temperature coherent control of defect spin qubits in silicon carbide. *Nature.* **479**, 84–87 (2011).

54. Shang, Z,. *et al.* Microwave-assisted spectroscopy of vacancy-related spin centres in hexagonal SiC. *Phys. Rev. Applied.* **15**, 034059 (2021).

55. Fanciulli, M., Electron paramagnetic resonance and relaxation in BN and BN:C, *Philosophical Magazine*, **76**, 3 (1997).

56. Katzir, A., Suss, J. T., Zunger, A. & Halperin, A., Point defects in hexagonal boron nitride. I. EPR, thermoluminescence, and thermally-stimulated-current measurements. *Phys. Rev. B.*, **11**, 6 (1975).

57. Yang, T.C. *et al.* Zero-field magnetic resonance of the photo-excited triplet state of pentacene at room temperature. *J. Chem. Phys.* **113**, 11194 (2000).

Supplementary Information for

# Room-temperature optically detected magnetic resonance of single defects in hexagonal boron nitride

Hannah L. Stern[1*†], John Jarman[1*], Qiushi Gu[1], Simone Eizagirre Barker[1], Noah Mendelson[2], Dipankar Chugh[3], Sam Schott[1], Hoe H. Tan[3], Henning Sirringhaus[1], Igor Aharonovich[2] and Mete Atatüre[1†].

[1] *Cavendish Laboratory, University of Cambridge, J.J. Thomson Ave., Cambridge, CB3 0HE, United Kingdom.*

[2] *ARC Centre of Excellence for Transformative Meta-Optical Systems, Faculty of Science, University of Technology Sydney, New South Wales, Australia.*

[3] *ARC Centre of Excellence for Transformative Meta-Optical Systems, Research School of Physics and Engineering, The Australian National University, Canberra, Australian Capital Territory, Australia.*

## Contents

1. **Material characterisation**
    I. Confocal measurements
    II. Laser power saturation of integrated-PL intensity
    III. Second-order autocorrelation measurements
2. **ODMR of single defects**
    I. Determination of saturation conditions: microwave power
    II. Determination of saturation conditions: laser power
    III. Analysis of functions fit to model ODMR lineshape
    IV. Additional ODMR spectra of single defects
    V. PL spectra variation of single defects
3. **ODMR of high-density ensemble**
    I. g-factor calibration
    II. ODMR lineshape of high-density ensemble
4. **EPR of high-density ensemble**

# 1. Material characterisation

## I. Confocal measurements

Confocal measurements were performed on a room-temperature home-built confocal setup described in Methods. Figure S1a shows an example 8x6 µm confocal image obtained by integrated-PL intensity under 100-$\mu$W 532-nm laser excitation. Panel b is a high-resolution 1.5x1.5 µm confocal image of defect A.

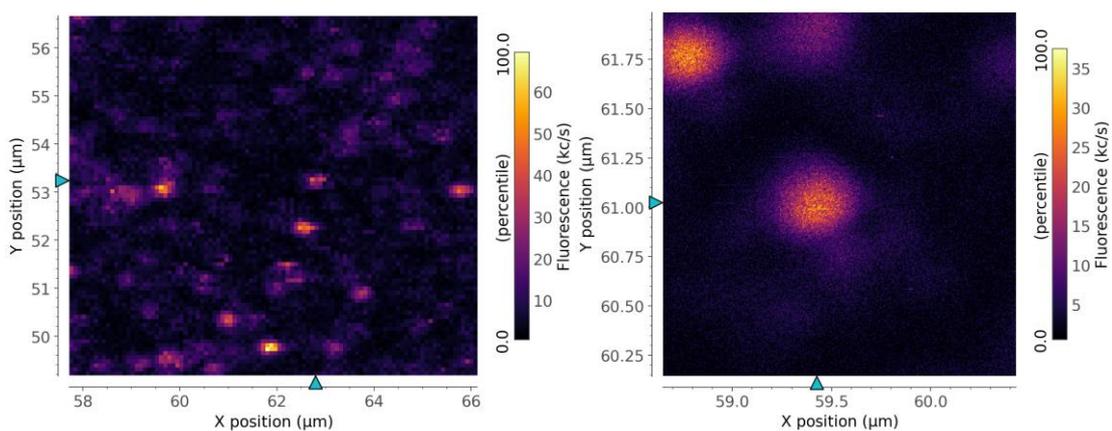

Figure S1: Confocal maps of the low-density hBN sample. (left) large-area image displaying multiple isolated defects, (right) high-resolution image of defect A.

## II. Laser power saturation of integrated-PL intensity

Photoluminescence saturation measurements as a function of laser power for emitters A-C were recorded with 532-nm excitation up to 350-$\mu$W power and the PL was collected through a 550-nm long pass filter.

Data is fit to the function:

$$I = \frac{I_{\text{sat}} P}{P + P_{\text{sat}}} \quad (1)$$

Where $I$ is the PL intensity, $P$ is the laser power, $I_{\text{sat}}$ is the PL intensity (cps) at saturation and $P_{\text{sat}}$ is the saturation laser power. We find that each defect shows clear saturation behaviour, and we note that there is significant variation in these parameters across the defects we measure.

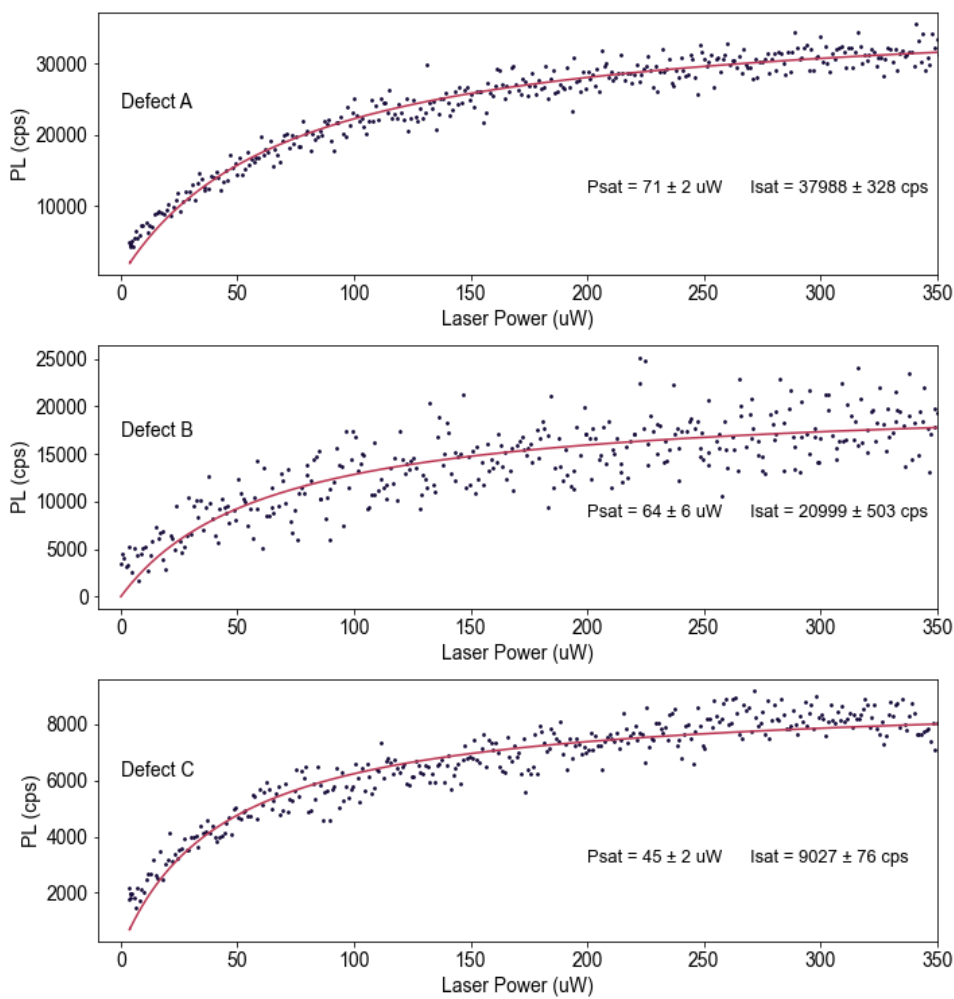

Figure S2: Integrated PL intensity as a function of the excitation laser power for defects A, B and C, presented in panels top, middle and bottom, respectively.

## III. Second-order autocorrelation, $g^{(2)}(\tau)$, measurements.

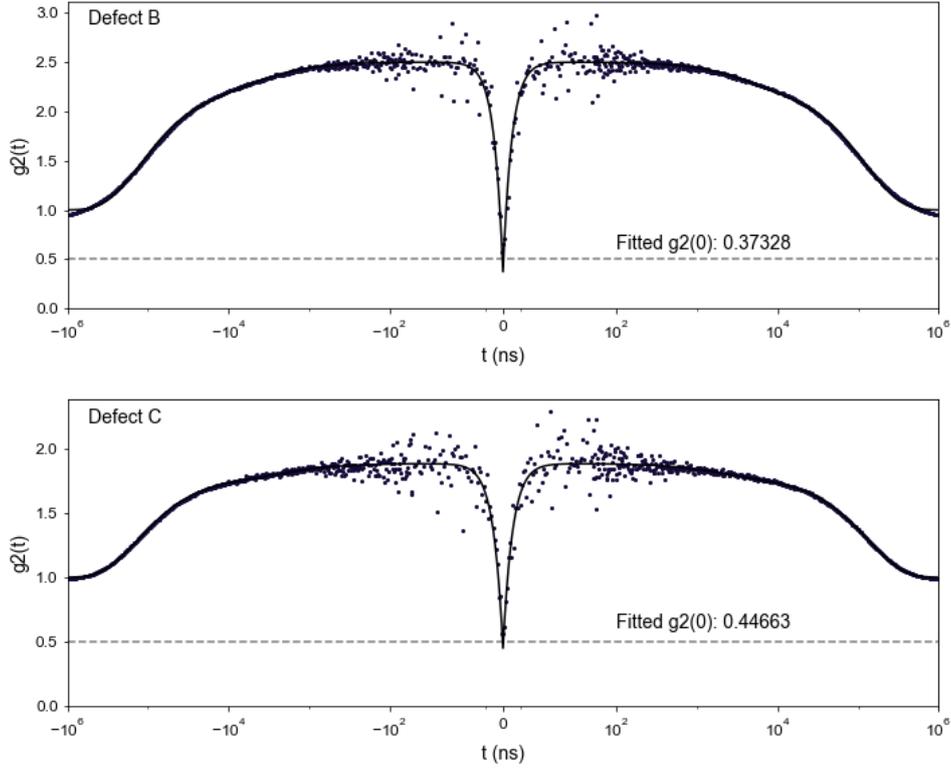

Figure S3: Second-order autocorrelation, $g^{(2)}(\tau)$, measurements for Defects B and C fit to triexponential function, given by Eq. 2. Measurements carried out at $1.6 P_{sat}^{optical}$ laser power for defect B and $2.1 P_{sat}^{optical}$ for defect C.

Fit functions used for $g^{(2)}(\tau)$ measurements:

$$g^{(2)}(\tau) = y_0 - ae^{(\tau-t0/\tau 1)} + be^{(\tau-t0/\tau 2)} + ce^{(\tau-t0/\tau 3)} \qquad (2)$$

|  | Defect A | Defect B | Defect C |
|---|---|---|---|
| $y_0$ | 36.5 ± 0.1 | 68.3 ± 0.4 | 21 ± 0.1 |
| a | 34.4 ± 1.4 | 145 ± 2 | 130 ± 0.6 |
| $\tau 1$ (ns) | 1.6 ± 0.1 | 4.9 ± 0.1 | 5.4 ± 0.2 |
| b | 5.0 ± 0.4 | 16 ± 0.6 | 2 ± 0.1 |
| $\tau 2$ (ns) | 2.5e4 ± 3.7e3 | 4.3e3 ± 375 | 1.9e3 ± 360 |
| c | 5.3 ± 0.5 | 86.2 ± 0.6 | 16 ± 0.2 |
| $\tau 3$ (ns) | 2.8e3 ± 383 | 1.2e5 ± 2.5e3 | 1.3e5 ± 3.7e3 |

Table S1: Fit parameters extracted from applying the above functions to the $g^{(2)}(\tau)$ data shown in Fig. 1 of the main text and Fig. S3 of the SI.

## 2. ODMR of single defects

### I. Determination of saturation conditions: microwave power

In the following analysis we show the microwave-power saturation ODMR measurements that were performed as a function of magnetic field in order to account for the frequency-dependent microwave transmission of our microstrip at each magnetic field.

We measure the ODMR with 2-MHz microwave frequency steps and present the data as an average of the two nearest data points. We then fit the ODMR signal to a single Lorentzian with the following function,

$$C = \frac{A\gamma^2}{(\nu-\nu_0)^2+\gamma^2} + y_0 \qquad (3)$$

where $C$ is ODMR contrast, $A$ is the amplitude, $\gamma$ is the half width half maximum, $\nu_0$ is the central frequency and $y_0$ is the y offset. To fit the ODMR spectra we first remove a static background of 0.2% which is a mechanical offset present in all our measurements.

Figure S4 shows the ODMR contrast determined from a single Lorentzian fit (Eq.3) to the ODMR central frequency, for defect A at a range of input microwave powers and magnetic field strengths, at 100-μW laser power. The data is fit to the saturation function (Eq1).

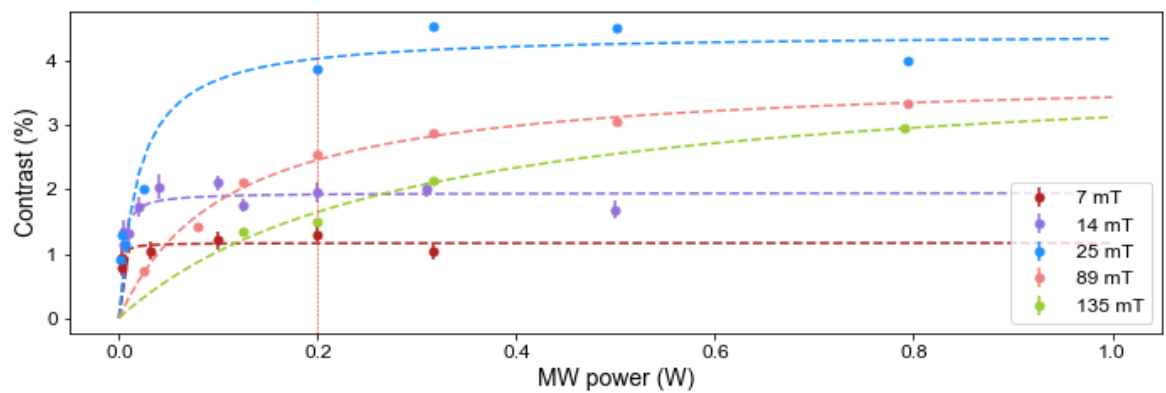

Figure S4: The ODMR contrast, at five different magnetic field strengths, as a function of input microwave power (W). The saturation fits (Eq1) give a saturation ODMR contrast of 1.2(1)%, 1.9(1)%, 4.4(3)%, 3.8(1)% and 4.0(4)% for 7 mT, 14 mT, 25 mT, 89 mT and 135 mT, respectively.

Fits to the data show that saturation occurs at different values of the microwave power for different microwave frequencies. At 135 mT 10$P_{sat}^{microwave}$ corresponds to 3-W input to the microstrip, a power at which we experience troublesome heating and drift. For this reason, we chose a fixed microwave power (0.2 W) for the measurements in Fig. 3.

## II. Determination of saturation conditions: laser power

The laser power dependence of ODMR contrast and linewidth is shown below for defect A, for the laser powers accessible to us (0-350 µW). The $P_{sat}^{laser}$ extracted from the fit is 75 µW. We use a laser power of 100 µW in the measurements in order to achieve the best signal-to-noise at the lowest optical drive possible. There is minimal linewidth broadening over the laser power range measured.

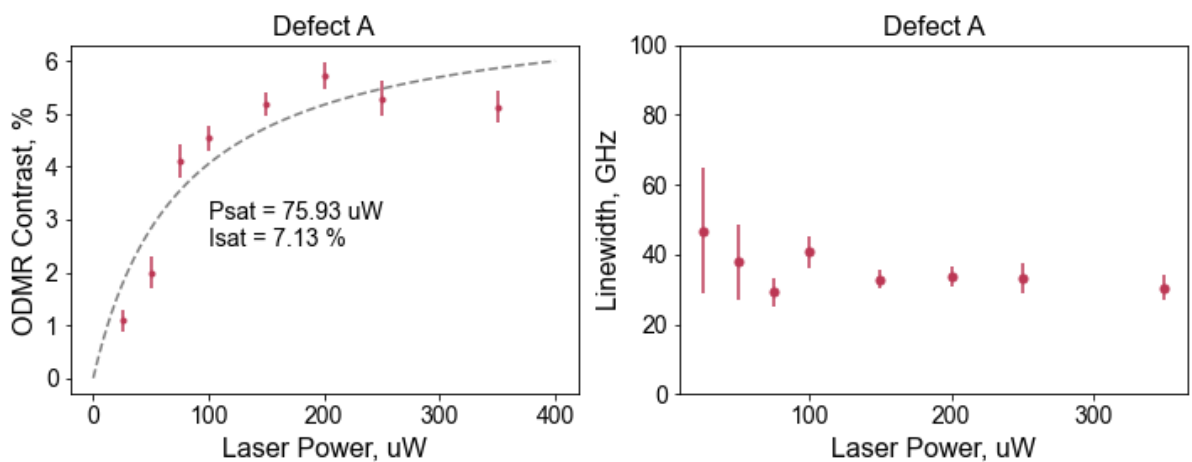

Figure S5: Laser power dependence the contrast and linewidth for defect A.

## III. Analysis of functions fit to model ODMR lineshape

Here we compare four distributions to model the lineshape of the ODMR spectra for defects A, B and C. The measurements were all recorded at $P_{sat}^{microwave}$ and an in-plane magnetic field.

For the Gaussian function we use the following equation,

$$C = A\, e^{(\frac{\nu - \nu_0}{2\sigma})^2} + y_0 \qquad (4)$$

where $A$ is the amplitude, $\nu_0$ is the central frequency, $\sigma$ is the variance and $y_0$ is the y offset.

For the Voigt function we use the following equation,

$$C = B \frac{1}{\sigma\sqrt{2\Pi}} e^{(\frac{\nu-\nu_0}{2\sigma})^2} + A \frac{\gamma^2}{(\nu-\nu_0)^2+\gamma^2} \qquad (5)$$

Where $C$ is the contrast, $B$ is the Gaussian amplitude, $\nu_0$ is the ODMR central frequency, $\sigma$ is the Gaussian variance and $A$ is the Lorentzian amplitude and $\gamma$ is the Lorentzian half width half max. For the Lorentzian we use Eq. 3 above. For the double Lorentzian we use two Lorentzians that are summed together to model the lineshape.

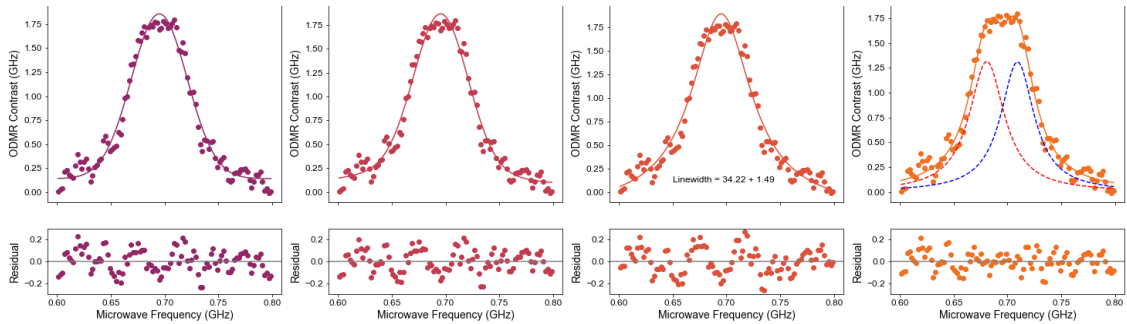

Figure S6: Defect A, 25 mT, $P_{sat}^{microwave}$ showing a double Lorentzian splitting of 29(2) MHz.

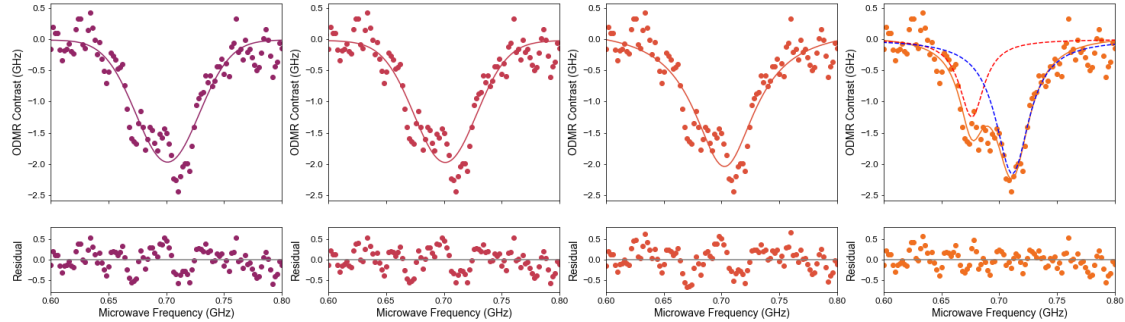

Figure S7: Defect B, 25 mT, showing a double Lorentzian splitting of 35(2) MHz.

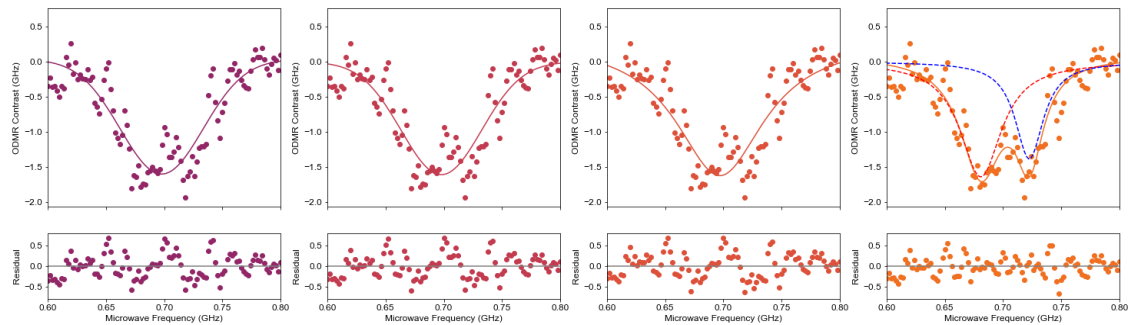

Figure S8: Defect C 25 mT, showing a double Lorentzian splitting of 42(3) MHz.

## IV. Additional ODMR spectra of single defects

From a total of 80 defects studied, we measure ODMR signal on nine defects. The survey was conducted at a constant in-plane magnetic field of 25 mT. The ODMR signatures of the defects are presented below in Fig. S9.

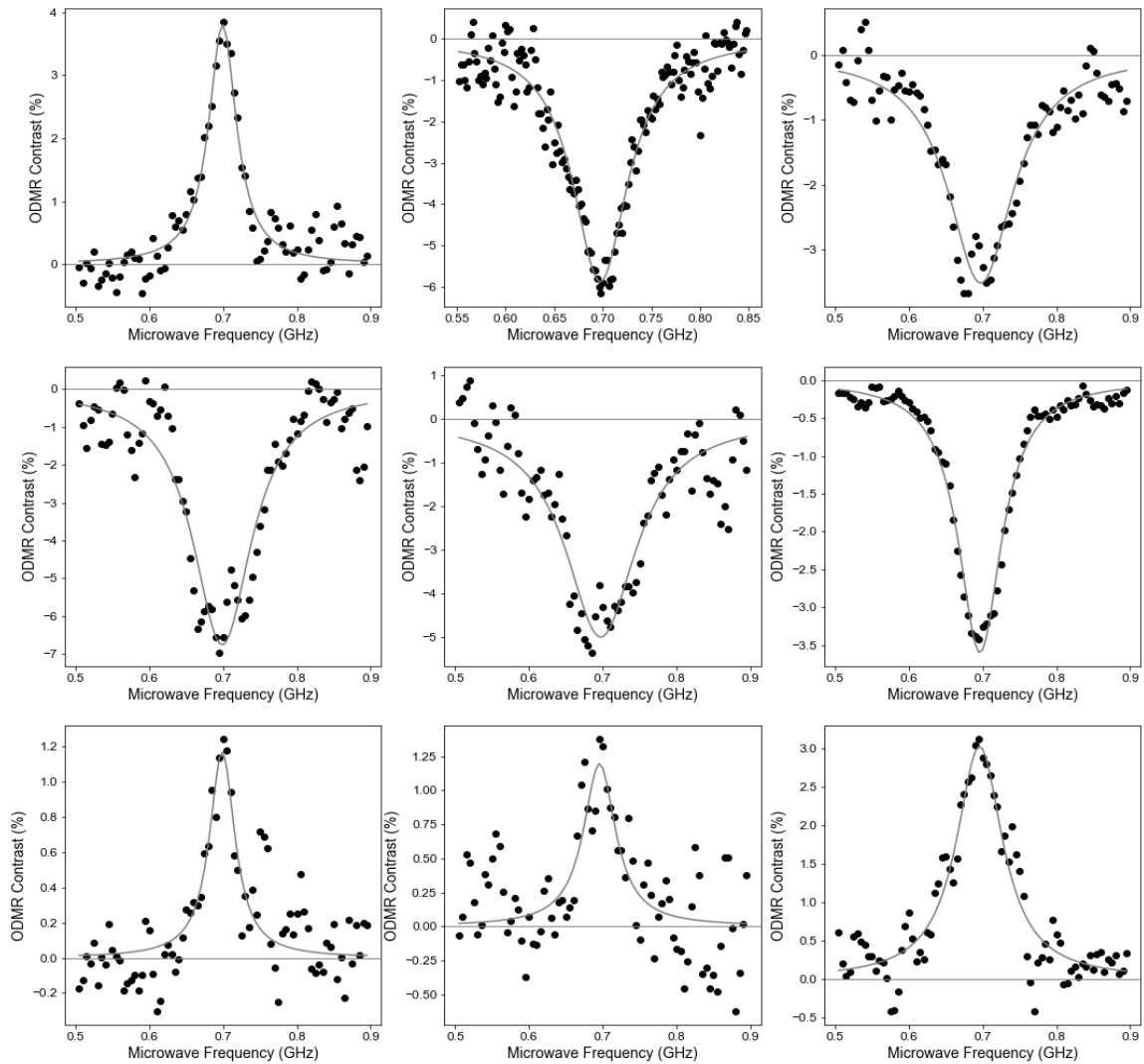

Figure S9: ODMR signal detected for nine defects at 25 mT at same optical and microwave excitation power, but with different data acquisition times.

## V. PL spectra variation of single defects

Photoluminescence spectra of 29 single-photon emitting defects in the hBN film were obtained under 532-nm CW illumination (see Methods). The variation in the zero-phonon line position between different defects is shown in Figure S13 below for ODMR inactive (grey circles) and ODMR active emitters (blue for those showing negative ODMR contrast, red for those showing positive ODMR contrast). For some defects, the emission profiles are clipped by the 550-nm long-pass filter placed to remove the laser light before detection, and therefore we can only estimate the position of their zero-phonon line (hollow circles). We find no clear correlation between the ZPL position and the presence or sign of ODMR contrast.

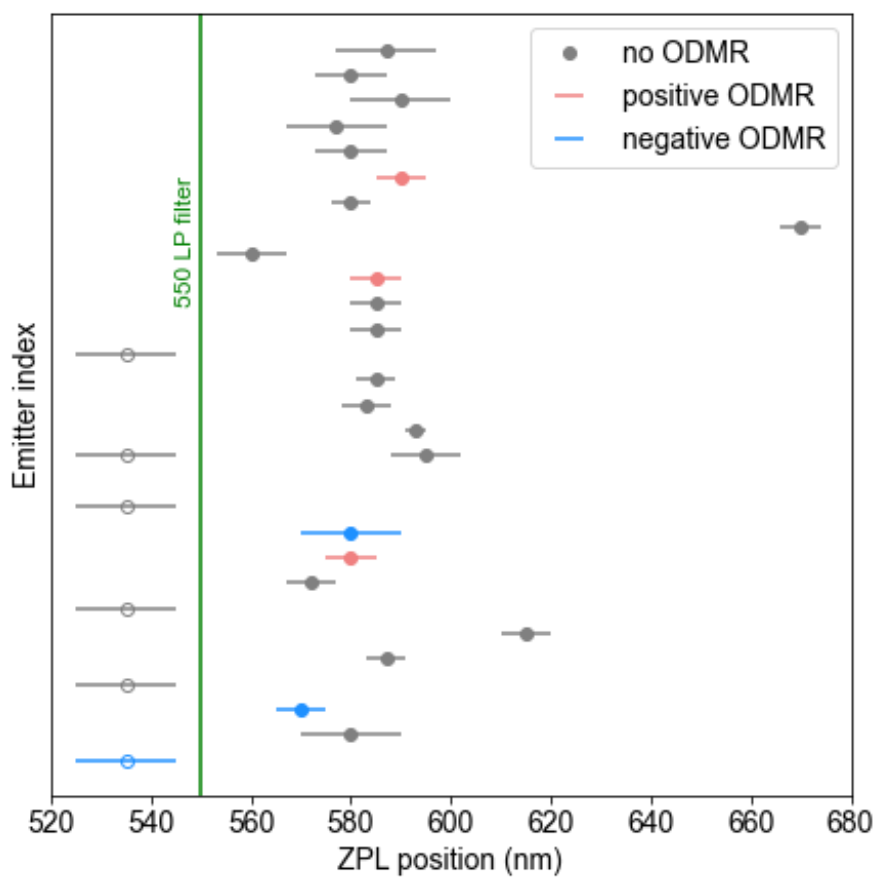

Figure S10: Variation in the zero-phonon line position for 29 single-photon emitting defects in the hBN film, as obtained from their photoluminescence spectra with 532-nm excitation.

## 3. ODMR of high-density ensemble

### I. g-factor calibration

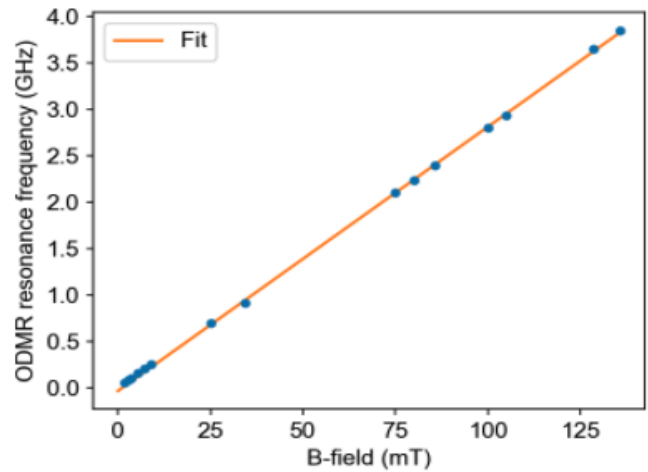

Figure S11: Frequency of central peak position against magnetic field strength of the ODMR resonance for the ensemble of spin defects. Each ODMR signal was fit to a Lorentzian (Eq. 3) to extract the central frequency of the resonance. The measurements were conducted with 532 nm laser excitation at 200 µW. A MW power of between 0.9-1 W was used across a loop antenna placed between the hBN film and the microscope objective.

### II. ODMR lineshape of high-density ensemble

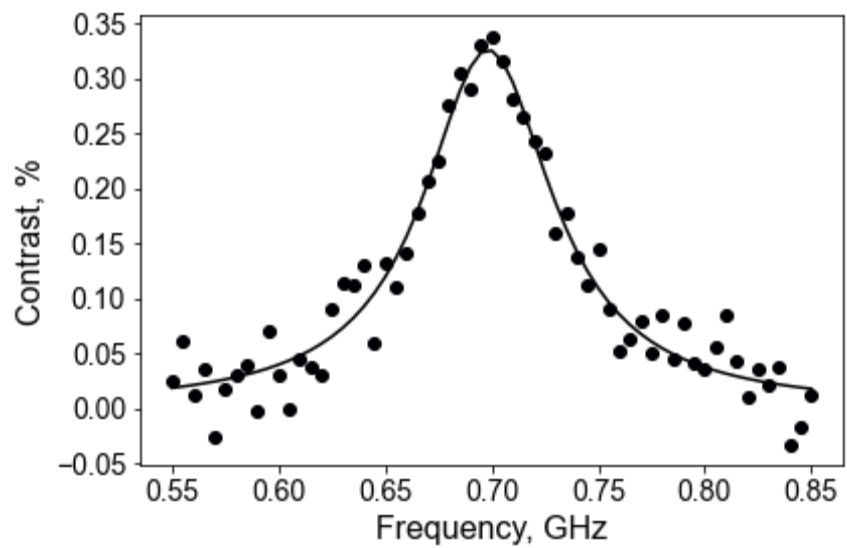

Figure S12: CW-ODMR spectrum of the ensemble of spin defects in the h-BN film fit to a Lorentzian. The measurement was conducted at 100 µW laser power and $P_{sat}^{microwave}$ microwave power applied across the microwave microstrip (see detailed in Methods in the main text). The linewidth determined from a single Lorentzian fit is 37(2) MHz.

## 4. EPR of high-density ensemble

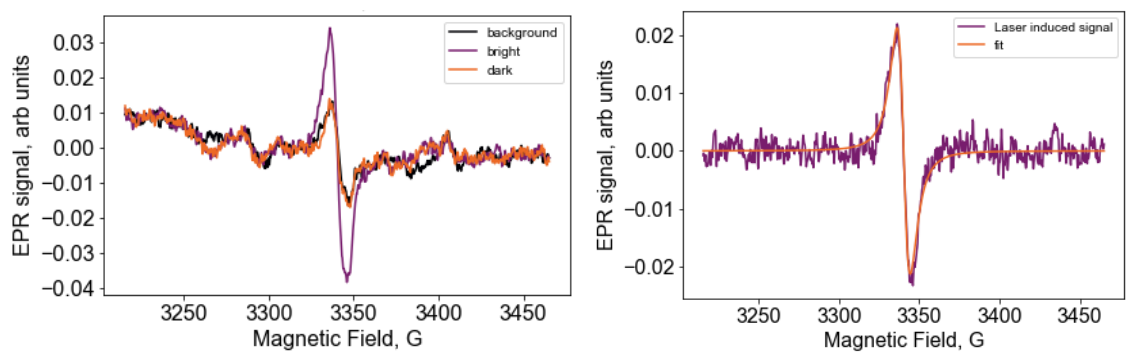

Figure S13: (left panel) The room temperature background, dark and light-induced EPR signal for the ensemble, measured using 532 nm laser excitation with a power of 55 mW over an area of 2.5 mm diameter spot on sample. (right panel) The light-induced EPR signal for the ensemble with background and dark signals subtracted. A Lorentzian fit to the linewidth gives 7.1 G (~20 MHz).